\documentclass[doublecol]{epl2}
\usepackage{graphicx}
\usepackage{url}

\newcommand{\bk}{\ensuremath{{\bf k}}}
\newcommand{\bq}{\ensuremath{{\bf q}}}
\newcommand{\bp}{\ensuremath{{\bf p}}}

\newcommand{\intk}{\ensuremath{{\int\frac{d^3\bk}{(2\pi)^3}}}}

\newcommand{\intp}{\ensuremath{{\int\frac{d^3\bp}{(2\pi)^3}}}}

\title{Mode-coupling as a Landau theory of the glass transition}
\shorttitle{MCT as a Landau Theory} %Insert here a short version of the title if it exceeds 70 characters
\author{A. Andreanov\inst{1}, G. Biroli\inst{2} and J.-P. Bouchaud\inst{3}}
\shortauthor{A. Andreanov, G. Biroli and J.-P. Bouchaud}
\institute{
	\inst{1} Theoretical Physics, University of Oxford, 1 Keble Road, Oxford, OX1 3NP, United Kingdom\\
	\inst{2} Institut de Physique Th\'eorique, CEA, IPhT, F-91191 Gif-sur-Yvette, France and CNRS, URA 2306\\
	\inst{3} Science and Finance, Capital Fund Management, 6 Bd Haussmann, 75009 Paris, France\\ 
}

\abstract{
We derive the Mode Coupling Theory (MCT) of the glass transition as a Landau theory, formulated as an expansion of the exact dynamical equations in the difference between the correlation function and its plateau value. This sheds light on the universality of MCT predictions. While our expansion generates higher order non-local corrections that modify the standard MCT equations, we find that the square root singularity of the order parameter, the scaling function in the $\beta$ regime and the functional relation between the exponents defining the $\alpha$ and $\beta$ timescales are universal and left intact by these corrections.
}

\begin{document}

\maketitle

The Mode-Coupling Theory of glasses (MCT), developed since the mid-eighties following the seminal work of G{\"o}tze \cite{BGS} and Leutheusser \cite{Leuthesser}, has significantly contributed to our understanding of the slowing down of supercooled liquids. One of its cardinal predictions is the appearance of a non trivial $\beta$-relaxation regime where dynamical correlation functions pause around a plateau value before finally relaxing to zero. In the vicinity of this plateau value, the theory predicts two power-law regimes in time (or in frequency), and the divergence of two distinct relaxation times, $\tau_\alpha$ and $\tau_\beta$, at the MCT critical temperature $T_d$. Although this divergence is smeared out by activated events in real liquids, the two-step relaxation picture suggested by MCT seems to account quite well for experimental and numerical observations  \cite{MCTreviewdas}, at least in weakly supercooled liquids and for hard sphere colloidal systems.

Originally, MCT was obtained as an uncontrolled self-consistent approximation within the Mori-Zwanzig projection operator formalism for Newtonian particles. This scheme yields an integro-differential equation for the dynamic structure factor $C({\bf k},t)$ that captures mathematically the slowing down of the dynamics and the appearance of a two-step relaxation at equilibrium. It provides very detailed predictions for the scaling properties of $C({\bf k},t)$ in the vicinity of the plateau value $f_{{\bf k}}S_{{\bf k}}$, where $S_{{\bf k}}$ is the static structure factor and $f_{{\bf k}}$ is called the non-ergodicity parameter (akin to the Edwards-Anderson parameter in spin glasses). Alternative derivations of MCT based on field theory have been sought for \cite{DasMazenkoprl} and research on this topic has continued until now \cite{reichman-fdt,abl,KimKawasaki,Hayakawa}. It was also realized  that the same integro-differential equations describe the exact evolution of the correlation function of mean-field p-spin glasses \cite{reviewBCKM}. This is important for at least two reasons:
\begin{itemize}
\item Technically, it shows that the MCT approximation is {\it realizable}: there is a well defined system for which it is exact; hence MCT does not violate basic physical constraints.
\item Physically, it brings in a very useful interpretation of the MCT freezing transition in the ``energy landscape'' parlance. Above the transition $T_d$, the dynamics is dominated by unstable saddles that become progressively less and less unstable as one approaches the transition; below the transition, there are only local minima that are separated by infinite barriers in mean-field, so that the system is forever trapped in one of them \cite{reviewBCKM}. For non mean-field systems, these barriers are finite and the transition is smoothed. MCT can be naturally embedded within the broader Random First Order Theory \cite{reviewRFOT} of the glass transition. In this context it describes the high temperature region where metastable states are still in embryo.
\end{itemize}
However, this analogy shows that MCT is (at best) an incomplete theory of real supercooled liquids,  and needs to be corrected and enhanced. An important question is to know to what extent the quantitative predictions of MCT are stable with respect to the ignored contributions. In fact, it is genrally accepted that the quantitative value of the critical temperature (or the critical density) obtained solving MCT equations is incorrect, as is the 
interaction parameter $\lambda$, which is a functional of the static structure factor and fixes the value of critical exponents. However, when MCT is used to fit empirical data, it is assumed (without much justification) that the predictions about the critical behavior remain correct if $T_d$ and $\lambda$ are treated as adjustable parameters. This implicitly assumes that some MCT predictions are universal, e.g. the square root singularity of $f_{{\bf k}}$ and the relation between the exponents describing the divergence of $\tau_\alpha$ and $\tau_\beta$, and others which are not, e.g. $T_d$ and the actual value of the exponents!

Clearly, the present theoretical understanding of MCT needs to be improved. Assessing the degree of {\it structural stability} of the theory and its universality properties is a crucial issue to resolve both for theoretical and practical purposes. One step in this direction has been performed by Szamel \cite{szamelbeyondmct} and then later generalized by Mayer et al., where a schematic version of MCT including higher order correlations was proposed and analyzed  \cite{reichman-gmct}. The result is that whenever the theory is truncated at any finite order in the n-boby correlations, the phenomenology of MCT is exactly recovered with a finite $T_d$,  whereas $T_d=0$ when the theory is treated exactly to all orders. Another approach was followed in \cite{abl}, based on a field-theoretical formulation of MCT consistent with the Fluctuation-Dissipation Theorem, which suggests closure schemes different from standard MCT.

The aim of this work is to argue that MCT can be rephrased as a Landau theory of the glass transition, based on general assumptions about the nature of the dynamical arrest but {\it without relying on any particular model}.  Therefore, some predictions of MCT are indeed generic and should be useful in a certain regime of time and temperature. Note that a Landau theory for the glass transition has been developed also in \cite{bulbul}, but it has a very different starting point and it does not focus on MCT.

The Landau theory is a general phenomenological approach to equilibrium phase transitions~\cite{landau,toledano}. It relies on a number of natural hypotheses, such as symmetry, genericity and regularity. In the classic example of the Ising model for ferromagnets, the expansion of the free energy as a function of the magnetisation $m$ reads, in the homogeneous case and subject to external field $h$: $F[m]=F_0 - mh + \frac{b}{2}(T-T_c)m^2+\frac{g}{4!}m^4+...$, from which a certain number of well known mean-field properties can be derived. Including the first gradient corrections in the inhomogeneous case also allows one to show that close to $T_c$, the divergence of the uniform susceptibility $\chi(\bq=0)$ is accompanied by the divergence of the correlation length $\xi$, over which magnetisation fluctuations are correlated. The Landau construction can falter in three distinct ways:
\begin{enumerate}
\item  Higher order terms, neglected in the expansion of $F[m]$ as a series of $m$, could qualitatively change the above predictions (structural instability). This happens, for example, close to a multicritical point where $g(T_c)=0$. But if the transition remains second order, higher order terms are truly negligible when $\epsilon=|T-T_c| \to 0$ and the predictions are universal.
\item  The non-linear feedback of spatial fluctuations on the divergence of the susceptibility can change all the critical exponents when the dimension of space is smaller than $d_u=4$ in the case of the Ising model. For $d > d_u$, on the other hand, one can prove that the low-$q$ behaviour of $\chi(\bq)$ is (close to the critical point) identical to that predicted by Landau's theory. 
\item Non-perturbative effects can wipe out the transition. This is the case for example of the spinodal transition: the system is not able to reach the critical point because of nucleation, which is an activated process. 
\end{enumerate}
Even the analogue of point (1) is difficult in the case of MCT; the basic reason being that the order parameter is a not a scalar, but it is a time dependent function  $C(\bk,t)$. The proof that MCT is structurally stable with respect to the addition of higher order terms is already quite complex and this will be the scope of the present paper. Once this is achieved, one should still worry about points (2) and (3) above. As already mentioned, it was recently realized that diverging fluctuations and an upper critical dimension $d_u$ also exist for MCT~\cite{bb,bbmr,stokes-einstein} (see also \cite{KT,FP} for earlier insights). In order to complete our proof that MCT is stable, one should prove that spatial fluctuations can be safely neglected in $d > d_u$ and understand how close one can get to the critical point before non-perturbative (activated) effects impair the transition.  We will completely disregard these issues in the present paper, and  focus only on point (1).

\begin{figure}

\includegraphics[scale=0.3, angle=270]{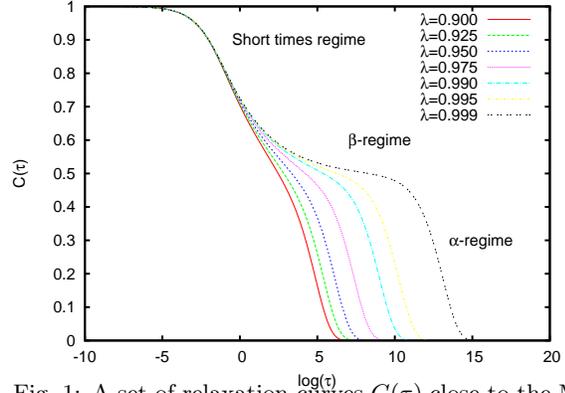}
\vspace{1mm}
\caption{\label{two-step}A set of relaxation curves $C(\tau)$ close to the MCT transition for the so called schematic model of MCT \cite{MCTreviewdas} and several values of $\lambda$, which quantifies the deviation from the transition. As $\lambda\to 1$ a shoulder develops in the relaxation pattern corresponding to the emerging $\beta$-relaxation regime.}
\end{figure}

The case of the glass transition is quite different from standard critical phenomena. Several physical and formal problems prevent a direct analogy. The glass transition seems to be a purely dynamical phenomenon: simple static, thermodynamical properties do not present any peculiarities as the liquid freezes into a glass \footnote{It has however been argued that highly non-trivial static correlations, called point to set correlations, increase approaching the glass transition \cite{pointtoset}.}. The above Landau construction simply does not make sense in the absence of the clear analogy of the free energy. This means that the order parameter in glasses cannot be a one point function (such as the magnetisation) but, instead, it is likely to be a two point dynamic correlation. The slowing down of the dynamics in glasses is found to be accompanied by the appearance of a plateau value $f_\bk$ in the relaxation pattern of the dynamical structure factor $C(\bk,\tau)$. Since the appearance of a plateau coincides with increasing time scales, one expects that within a very long time interval (to be specified), the correlation function can be approximately written as:
\begin{equation}
C(\bk,\tau) \approx f_\bk S_\bk + \delta C(\bk,\tau), \qquad   \delta C(\bk,\tau) \ll 1.
\end{equation}
The idea underlying our construction of a Landau theory for glasses is to consider $\delta C(\bk,\tau)$ as the analogue of the order parameter $m$ and construct a general, structurally stable, dynamical equation for  $\delta C(\bk,\tau)$. A way to construct such an equation is to start from the exact dynamical evolution for $C$ and the response function $R$ that can be derived in the framework of various dynamical field theories, for example based on Dean's equation~\cite{dean} for Brownian dynamics, on Fluctuating Hydrodynamics for Newtonian dynamics \cite{DasMazenkoprl} (see also \cite{reichman-fdt,abl,KimKawasaki,Hayakawa}), or on Langevin equations for $p$-spin models \cite{reviewBCKM}. Although these theories give very different sets of equations, they can all be reduced to the following single equation in the ergodic region:
\begin{equation}
\label{gt-main}
\partial_\tau C(\bk,\tau)+T C(\bk,\tau)+\int\limits_0^\tau du\Sigma(\bk,\tau-u)\partial_u C(\bk,u)=0
\end{equation}
with initial condition $C(\bk,0)=S_\bk$. The self-energy $\Sigma(\bk,\tau)$ (or memory kernel in MCT terminology) is given by the sum of $2$-particle irreducible ($2$-PI) diagrams built with $C$ and $R$ lines, see e.g. \cite{abl}. We do not specify the details of the field theory underlying this equation, nor the Feynman rules for the diagrams contributing to $\Sigma$: we just need that such a theory exists. We also stay in the high-$T$ region, so that the system is at equilibrium: both $C$ and $R$ are then time translation invariant and the Fluctuation-Dissipation theorem holds at a diagrammatic level: $T \, R(\bk,\tau)=-\partial_\tau C(\bk,\tau)$. In this case the self-energy is a functional of the correlation function only. Eq. (\ref{gt-main}) has exactly the structure of the standard MCT equation for liquids, although there is no well defined prescription to build a consistent approximation for $\Sigma$; in this sense MCT is rather arbitrary and difficult to improve upon in a systematic way. The standard MCT results correponds to a self-consistent $1$-loop approximation for $\Sigma$, $\Sigma(\bk,t-s)=\intk V(\bk,\bp)C(\bk-\bp,t-s)C(\bp,t-s)$, where $V(\bk,\bp)$ is an effective vertex. \footnote{Note however that there are complications related to fluctuation dissipation relations \cite{reichman-fdt,abl,KimKawasaki,Hayakawa}.} When generalized to higher order diagrams, an important difficulty emerges: the non-locality in time of the corrections. Our goal is to prove that the main results of the standard MCT (or $1$-loop) approximation still hold. The proof is done as for usual  theories. We start from some conjectures about the critical properties, such as the nature of the order parameter and its critical properties, that are motivated by experimental and numerical findings. Then we show that they result from a Landau-like expansion, which allows one to assess their universal character and to fix the value of the critical exponents. We therefore assume that the order parameter is the dynamical correlation function and that displays the following features:
\begin{itemize}
\item There is structural arrest: below some temperature $T_d$, $\lim\limits_{\tau\to\infty}C(\bk,\tau) = f_\bk S_\bk$ with $f_\bk >  0$.
\item When $\epsilon=(T-T_d)/T_d \ll 1$, the correlation function exhibits a two-step pattern with three well separated characteristic time scales, see Fig~\ref{two-step}. We assume that there exists a diverging time scale $\tau_\beta(\epsilon)$ where the difference $\delta C(\bk,\tau)$ between $C$ and the plateau value is small, of order $r(\epsilon) \ll 1$. More precisely, the correlation function decay is decomposed into: (i) a short time regime, $\tau\sim \tau_0$, where $C(\bk,\tau)=C_0(\bk,\tau)$, with $C_0(\bk,\tau \gg 1) \to f_\bk S_\bk$: (ii) a $\beta$-regime, $\tau=s\tau_\beta(\epsilon)$ with $s=O(1)$: $\delta C(\bk,s\tau_\beta)=r(\epsilon) S_\bk (1-f_\bk)^2 G(\bk,s)$; (iii) an $\alpha$-regime, $\tau=s'\tau_\alpha(\epsilon)$ with $s'=O(1)$: $C(\bk,s'\tau_\alpha)=C_\alpha(\bk,s')$ describes the final fall off of the relaxation.
\end{itemize}

We assume further (and justify later) that the function $G(\bk,s)$ can itself be expanded in powers of $r(\epsilon)$: $G(\bk,s)=\sum_{n=1}^\infty r^{n-1} G_n(\bk,s)$. \footnote{One could have also some regular (in $\epsilon$) contributions. However, as we shall show the only two possible values of $r(\epsilon)$ are $\sqrt{\epsilon}$ or $\epsilon$. As a a consequence, regular contributions will be automatically contained in the expansion.} All functions $G_n$ are a priori singular at $s=0$ and $s=\infty$, reflecting the fact that the behaviour of $C$ must match the short time regime and $\alpha$ regime, where the deviation from the plateau ceases to be small. A crucial remark for the following is that any function $G_n$ will appear with a prefactor $r^n(\epsilon)$. These hypotheses turn out to be sufficient to generalize the MCT results in the $\beta$-regime. First, it is clear that the above expansion of $G(\bk,s)$ generates a similar expansion of the self-energy $\Sigma[C]$ in the $\beta$-regime: $\Sigma(\bk,s)=\sum_{n=1}^\infty r^{n-1} \Sigma_n(\bk,s)$, where the $\Sigma_n$ do not depend on $\epsilon$, but are some functionals of $C(\bp,\tau)$. The most generic functional form for $\Sigma$ {\it a priori} includes contributions from all three regimes:
\begin{equation}
\Sigma(\bk,s)=\Sigma[\{C_0(\bp,s'\tau_\beta),r(\epsilon)G(\bp,s'),C_\alpha(\bp,s'\tau_\beta/\tau_\alpha)\}],
\end{equation}
but since the $\Sigma_n$ should not depend on $\epsilon$, general arguments can be used to restrict the actual functional form of $\Sigma_n$. Note that we have used the notation $s'$  to stress that this equation is a functional relation which is non-local in time. Also, even if we had assumed that $\delta C(\bk,s\tau_\beta)$ only contains a single term of the order of $r(\epsilon)$ then we would have generated corrections of all orders in $r(\epsilon)$ anyway. The reason is that the self-energy will contain all orders in $r(\epsilon)$ as it can be found by expanding the above equation to all order in $r(\epsilon)G(\bp,s)$; this will feed back, via the Schwinger-Dyson equations, on $G(\bp,s)$ itself.

We now illustrate how this works for the lowest order terms  $\Sigma_0$, $\Sigma_1$ and $\Sigma_2$. As we shall see higher orders are in fact irrelevant for our purpose. Clearly, the zeroth order term  $\Sigma_0(\bk,s)$ can only be a function of the wavevector $\bk$ since in the limit $\epsilon \to 0$ time scales separate: $\tau_\beta\to\infty$ and $\tau_\beta/\tau_\alpha\to 0$ and therefore the previous equation 
implies that in $\Sigma_0(\bk,s)$ all dependence on $s$ drops out. The first order contribution $\Sigma_1(\bk,s)$ must read:
\begin{equation}
\Sigma_1(\bk,s) = \int\limits_0^\infty du\int_{\bp} K_1(\bk,\bp;s,u) G_1(\bp,u) \quad
\end{equation}  
where, henceforth, we shall use the notation $\int_{\bp}=\intp$. Any other combination containing some $G_n$ gives an extra factor $r^n(\epsilon)$ and thus it corresponds to a higher order contribution. In the original time variables, the kernel $K_1$ must have some regular shape with a span fixed by the microscopic time scale. Therefore, in the rescaled variables $u,s$, $K_1$ must be local in $u-s$ and, to lowest order, is a $\delta$ function; higher derivatives of the $\delta$ function correspond to corrections smaller by at least a factor $\tau_0/\tau_\beta$ which, as we shall see, turn out to be negligible even at order $r^2$. Therefore, $\Sigma_1(\bk,s) = \intp K_1(\bk,\bp) G_1(\bp,s)$.

The second order term has a richer structure. First, there is a term similar to the first order one with $G_2$ instead of $G_1$: $\intp K_{2}(\bk,\bp)G_2(\bp,s)$. But since the kernel $K_2$ is obtained, as $K_1$, from the first derivative of the self energy with respect to $r(\epsilon)G(\bp,s)$, one finds $K_2=K_1$. Second, there are terms quadratic in $G_1$:
$$\int_{\bk_1,\bk_2}\int\limits_0^\infty du\int\limits_0^\infty dv K_{11}(\bq,\bk_1,\bk_2;s,u,v)G_1(\bk_1,u)G_1(\bk_2,v)$$ 
For the same reasons outlined above, time dependence of $K_{11}(\bq,\bk_1,\bk_2;s,u,v)$ is composed of $\delta$-functions and their derivatives. Some thinking about the underlying diagrammatic structure of the theory allows one to be convinced that the general structure of $K_{11}$ is, to leading order:
\begin{eqnarray}
K_{11}(s,u,v) = K_{11,\ell}\delta(s-u)\delta(s-v)+\hspace{2.9cm}&&\nonumber\\
K_{11,n\ell}\delta(u+v-s)(\partial_u+\partial_v)+\tilde K_{11,n\ell}(\partial_u+\partial_v)\delta(u+v-s)&&\nonumber
\end{eqnarray}
where to simplify the notation we have dropped all wave-vector dependence in the above equation. The fact that only the combination $u+v-s$ enters comes from causality and the separation of time scales. The full justification of the above form and other technical details\footnote{In particular one could think that the separation of timescales not only leads to delta function terms but also to constants. The latter are absent. This can be shown using the general field theoretical expression of the self energy, see \cite{alexei}.} are presented in \cite{alexei}. The first local term (in $s$), $K_{11,\ell}$, is like the usual MCT contribution, but the other two terms do not appear within standard MCT. The third term actually reduces to the second one plus local terms via integration by parts. 

This expansion is the main result in the construction of the Landau theory. The zeroth order equation in $r(\epsilon)$ fixes the non-ergodic parameter such that $\frac{T_d f_\bk}{1-f_\bk}=\Sigma_0(\bk)$, as in standard MCT. Substituting the expansion of $\Sigma$ up to second order in $r$ into (\ref{gt-main}) and using the expansion of $C$ in the $\beta$ timescale, one finally obtains, for $T=T_d(1+\epsilon)$ and in Laplace space within the $\beta$-regime (we have dropped zeroth order, as discussed above):
\begin{eqnarray}
&&r\left[T_d(1+\epsilon)\hat G_1(\bk,z)-\int_\bp K_1(\bk,\bp)\hat G_1(\bp,z)\right]+
\nonumber\\
& & r^2 \left[T_d(1+\epsilon)\hat G_2(\bk,z)-\int_\bp K_1(\bk,\bp)\hat G_2(\bp,z)\right]+\nonumber\\
& & \frac{T_d f_\bk\epsilon}{z(1-f_\bk)}
+ r^2 T_d(1+\epsilon)(1-f_\bk)z\hat G_1^2(\bk,z)=\nonumber\\
\label{gt-gen-eqn}
&&r^2\int_\bp K_{2}(\bk,\bp)\hat G_2(\bp,z)+\\
& + & r^2\int_{\bk_1}\int_{\bk_2} K_{11,\ell}(\bk,\bk_1,\bk_2) {\cal L}[G_1(\bk_1,\tau)G_1(\bk_2,\tau)](z)\nonumber\\
& + & r^2\int_{\bk_1}\int_{\bk_2} K_{11,n\ell}(\bk,\bk_1,\bk_2)z\hat G_1(\bk_1,z)\hat G_1(\bk_2,z),\nonumber
\end{eqnarray}
Identifying the coefficients order by order produces a series of equations. The first order fixes the yet unknown function $r(\epsilon)$: the expansion (\ref{gt-gen-eqn}) only contains terms with integer powers of $\epsilon$. They should be matched with powers of $r(\epsilon)$. Inspection of (\ref{gt-gen-eqn}) shows that there are two possibilities\footnote{Actually, there are other possibilities that correspond to higher order MCT singularities, which have been called $A_n$ \cite{gotze-sjogren}. In a usual Landau theory these correspond to tricritical, or even higher order, critical points. We will neglect them here since they require some fine tuning of the coupling constants.}: either $r=\epsilon$, or $r=\sqrt{\epsilon}$. The first choice yields a time independent solution for $G_1$ which is in contradiction with our hypothesis of a two-step relaxation with diverging time scales. Hence $r(\epsilon)=\sqrt{\epsilon}$, precisely as for usual MCT, or at $1$-loop order. This follows from the presence of a regular in $T$ term in (\ref{gt-main}).
%We expect this, in the spirit of a Landau theory, to be a general property of dynamical equations for glass-forming systems. 

The equation to order $r=\sqrt{\epsilon}$ now reads:
\begin{equation}
\label{gt-gen-1-eqn}
T_d\hat G_1(\bk,z)=\intp K_1(\bk,\bp)\hat G_1(\bp,z)
\end{equation}
This is the standard eigenvalue problem found within MCT that fixes the value of the critical temperature $T_d$. It constrains $G_1$ to be a product of wave-vector dependent and time dependent amplitudes, thus reproducing the well-know MCT ``factorization property": $\hat G_1(\bk,z)=\hat g(z)H_1(\bk)$, where $H_1$ is the right eigenvector of $K_1$ with largest eigenvalue $\Lambda=T_d$. At this order however the scaling function $\hat g(z)$ remains unfixed. The second order equation is trickier:
\begin{eqnarray}
\label{gt-gen-2-eqn}
&&T_d\hat G_2(\bk,z)-\int_\bk K_1(\bk,\bp)\hat G_2(\bp,z)=\nonumber\\
&&-\frac{T_d f_\bk}{z(1-f_\bk)} - T_d(1-f_\bk)z\hat g^2(z)H_1^2(\bk)+\\
&&+\int_{\bk_1}\int_{\bk_2} K_{11,\ell}(\bk,\bk_1,\bk_2){\cal L}[g^2](z) H_1(\bk_1)H_1(\bk_2)+\nonumber\\
&&+z\hat g^2(z)\int_{\bk_1}\int_{\bk_2}\hat K_{11,n\ell}(\bk,\bk_1,\bk_2)H_1(\bk_1)H_1(\bk_2)\nonumber
\end{eqnarray}
Following~\cite{gotze-1984}, we now multiply (\ref{gt-gen-2-eqn}) by $H_1(\bk)$ and integrate over $\bk$. The $G_2$ part of the equation vanishes and the remainder yields an equation on $\hat g(z)$. After some algebra and a proper rescaling of $z$ and $\hat g$ one finds:
\begin{equation}
\label{gt-gen-g-eqn}
\frac{1}{z}+\frac z \lambda \hat g^2(z)={\cal L}[g^2](z)
\end{equation}
where $\lambda$ is a constant that includes a non-local contribution as compared to MCT. But the structure of the equation on the scaling function $g$ is
exactly the same as in standard MCT. The properties of solution are well known: $g$ has a singular power law asymptotics at $z\to\infty$: $\hat g(z)\sim z^{a-1}$ and $z\to 0$: $\hat g(z)\sim z^{-1-b}$. The small time exponent $a$ and long time exponent $b$ characterize the decay of $\delta C(\bk,\tau)$ to and away from the plateau $f_\bk$. The exponents $a$ and $b$ are related by the famous equation:
\begin{equation}
\label{gt-gen-ab}
\frac{\Gamma^2(1-a)}{\Gamma(1-2a)}=\frac{\Gamma^2(1+b)}{\Gamma(1+2b)}=\lambda,
\end{equation}
which is a genuinely non-trivial and clear-cut prediction of MCT that constrains the range of values of $a$ and $b$ to: $0 \leq a < 1/2$ and $0 \leq b \leq 1$, in good agreement with experimental and numerical results. We have thus found that this relation has a much broader degree of validity and survives the introduction of an arbitrary number of loop corrections. The values of $a$ and $b$ are however different from the standard MCT (or $1$-loop) result, but as alluded to above, the parameter $\lambda$ is usually taken as an adjustable parameter anyway.

The fact that the form of the scaling function $g$ is the same as at $1$-loop (MCT) has two consequences. First, it fixes the functional dependence of the time scales exactly as for MCT:
\begin{equation}
\tau_\beta = \epsilon^{-1/2a}; \qquad \tau_\alpha=\epsilon^{-\gamma},
\end{equation}
$\gamma=1/2a+1/2b$. This is clear from the matching of $C(\bk,s\tau_\beta)$ at both ends of the $\beta$ regime. Second, it can be used to show that some superficial divergencies encountered in the calculation are in fact innocuous (see \cite{alexei} for more details). Note that the only extra contribution that appear in the generic case to second order in $g(s)$, namely the non-local term proportional to $\int du g'(u)g(s-u)$, does not modify the basic MCT equation, Eq. (\ref{gt-gen-g-eqn}).

The conclusion, which is the main result of this work, is that although $T_d$ and $\lambda$ are modified by taking into account corrections to MCT, the relation between the exponent $a$ and $b$,  the square root singularity as well as the scaling function $g$ are truly universal properties. This universality with respect to higher order {\it local} (in time) corrections was of course already shown by G\"otze long ago; here we have proven that this result is robust with respect to general {\it non-local} corrections as well, and suggesting that MCT has the status of a Landau theory of the glass transition~\footnote{As a consequence, it has a very different status compared to Mode-Coupling theories developed to compute critical exponents beyond mean field theory for standard critical phenomena.}.

The above schematic arguments can be made precise within the context of specialized model. We have in particular studied in full details the finite $N$ corrections to mean-field 3-spin glass model, where the structure of the perturbation theory can be used to check that the above conclusions hold in that case, see \cite{alexei} for details.

It was recently understood how MCT equations should be generalized in the presence of spatial inhomogeneities, where the correlation function $C$ can be space dependent: $C(\bk,\vec r;\tau)$, where $\vec r$ is the average of the two points $\vec r_1, \vec r_2$ between which the correlation is computed, and $\bk$ is the Fourier vector corresponding to $\vec r_1 - \vec r_2$. When wavelength of inhomogeneities is large, one can establish a gradient expansion of the MCT equations. In the schematic limit where all dependence on $\bk$ is discarded, the self-energy reads, to the lowest order:
$
\Sigma[C](s) = C(\vec r,s)^2 + w_1 C(\vec r,s) \nabla^2 C(\vec r,s) + w_2  \vec \nabla C(\vec r,s) \cdot \vec \nabla C(\vec r,s),
$
where $w_1$ and $w_2$ are some coefficients \cite{bbmr}. As mentioned in the introduction, these gradient terms are very important because they show how the MCT transition is in fact associated with a diverging correlation length, which corresponds to the scale over which a localized perturbation affects the surrounding dynamics \cite{bbmr}. The long-ranged critical fluctuations renormalize the value of the MCT exponents in $d < d_u=8$ \cite{stokes-einstein}. The above analysis, which was done in the homogeneous limit $\nabla \to 0$, should be repeated in the inhomogeneous case to complete our proof. We expect that the same conclusion will hold, namely that the results about dynamical correlation obtained within inhomogeneous MCT \cite{bbmr} are stable against the addition of higher order corrections. 

In conclusion, we have shown that MCT, which describes a specific slowing down mechanism through the progressive disappearance of unstable directions, has the status of a Landau theory and is therefore expected to make generic predictions, albeit polluted by activated events and critical fluctuations in finite dimensions. The interplay between critical fluctuations and activated events when $d < d_u$, and the crossover to low temperature dynamics is still largely an exciting open problem \cite{WolynesIndians}. Note also that even for the exact MCT equations, the critical region where the asymptotic scaling predictions are valid is unusually narrow \cite{Alba,ROMpaper}. It would 
be interesting to generalize our Landau approach to the aging regime and show 
what are the truly universal properties of the mean-field and MCT-like description of the aging dynamics \cite{reviewBCKM}.

In constructing the Landau theory, we have assumed that the freezing transition is discontinuous, with a finite value $f_\bk$ of the plateau at the transition. A viable alternative is of course that of a continuous transition of the spin-glass type, which leads to a completely different phenomenology. This raises the question of the possible realization of this second scenario in the context of supercooled liquids. All short-range interacting glasses seem to be characterized by rather small Lindemann parameters at the transition, meaning that it is hard to maintain any kind of amorphous long range order when individual molecules move substantially, and that the glass transition is therefore discontinuous \cite{WE}. This argument suggests that continuous glasses can only exist for long-ranged interacting particles or quantum systems. In the quantum case, it is imaginable that amorphous density waves can indeed form with a vanishing modulation amplitude (see \cite{VBG}). It would be very interesting to find experimental realizations of such a scenario.

\acknowledgments
We thank A. Lef\`evre for useful discussions. GB and JPB are supported by ANR Grant DYNHET; AA was supported in part by EPSRC Grant No. EP/D050952/1.


\begin{thebibliography}{80}
\bibitem{BGS} U. Bengtzelius, W. G{\"o}tze, A. Sj{\"o}ilander, \textit{J. Phys. C} \textbf{17}, 5915 (1984).
\bibitem{Leuthesser} E. Leuthesser, \textit{Phys. Rev. A} \textbf{29}, 2765 (1984).
\bibitem{MCTreviewdas}  S. P. Das, \textit{Rev. Mod. Phys.} \textbf{76}, 785 (2004).
\bibitem{DasMazenkoprl} S. P. Das, G. F. Mazenko, S. Ramaswamy, and J. J. Toner, \textit{Phys. Rev. Lett.} \textbf{54}, 118 (1985).
\bibitem{reichman-fdt} K. Miyazaki, D. R. Reichman, \textit{J. Phys. A: Math. Gen.} \textbf{38}, 20 (2005).
\bibitem{abl} A. Andreanov, G. Biroli, A. Lef\`evre, \textit{J. Stat. Mech.}, P07008 (2006).
\bibitem{KimKawasaki} B. Kim and K. Kawasaki, \textit{J. Stat. Mech.} (2008) P02004.
\bibitem{Hayakawa} T. H. Nishino and H. Hayakawa, \textit{Phys. Rev. E} \textbf{78}, 061502 (2008).
\bibitem{reviewBCKM} JP. Bouchaud,  L. Cugliandolo, J. Kurchan, M. M\'ezard, in {\it Spin glasses and Random Fields}, A.P. Young Editor (World Scientific) 1998.
%\bibitem{kob} W. Kob, in \textit{Slow relaxations and non-equilibrium dynamics in condensed matter}, vol. Session LXXVII of Les Houches Summer School (ed. J.-L. Barrat, M. Feigelman and J. Kurchan) published by EDP Sciences and Springer.
\bibitem{reviewRFOT} V. Lubchenko, P. G. Wolynes, \textit{Ann. Rev. Phys. Chem.} \textbf{58} 235 (2007).
\bibitem{szamelbeyondmct} G. Szamel, \textit{Phys. Rev. Lett.} \textbf{90}, 228301 (2003).
\bibitem{reichman-gmct} P. Mayer, K. Miyazaki, D. R. Reichman, \textit{Phys. Rev. Lett.} \textbf{97}, 095702 (2006).
\bibitem{bulbul} S. N. Majumdar, D. Das, J. Kondev, B. Chakraborty, \textit{Phys. Rev. E} \textbf{70}, 060501(R) (2004).
\bibitem{landau} L.D. Landau, \textit{Zh. Eksper. Teor. Fis.} \textbf{7}, 627 (1937).
\bibitem{toledano} J.-C. Tol\'edano, P. Tol\'edano, \textit{The Landau theory of phase transitions}, (World Scientific Publishing Co. Pte Ltd) 1987.
\bibitem{bb} G. Biroli, J.-P. Bouchaud, \textit{Europhys. Lett.} \textbf{67}, 21 (2004).
\bibitem{bbmr} G. Biroli, J.-P. Bouchaud, K. Miyazaki, D.R. Reichman, \textit{Phys. Rev. Lett.} \textbf{97}, 195701 (2006) .
\bibitem{stokes-einstein} G. Biroli, J.-P. Bouchaud, \textit{J. Phys.: Condens. Matter} \textbf{19} 205101 (2007).
\bibitem{KT} T. R. Kirkpatrick and D. Thirumalai,
Phys. Rev. A {\bf 37}, 4439 (1988).
\bibitem{FP} S. Franz and G. Parisi, 
J. Phys.: Condens. Matter {\bf 12}, 6335 (2000).
\bibitem{pointtoset}  J.-P. Bouchaud, G. Biroli, \textit{J. Chem. Phys.} \textbf{121}, 7347 (2004); 
G. Biroli, J.-P. Bouchaud, A. Cavagna, T. S. Grigera, P. Verrocchio, \textit{Nature Phys.} \textbf{4} 771 (2008);  M. M\'ezard, A. Montanari, J. Stat. Phys. {\bf 124} (2006) 1317. 
\bibitem{dean} D. S. Dean, \textit{J. Phys. A: Math. Gen.} \textbf{29}, L613 (1996).
\bibitem{alexei} A. Andreanov, Ph.D. thesis; \url{http://www.imprimerie.polytechnique.fr/Theses/Files/Andreanov.pdf}.
\bibitem{gotze-1984} W. G\"otze, \textit{Z. Phys. B - Condensed Matter} \textbf{59}, 195 (1985).
\bibitem{Alba} V. Krakoviack and C. Alba-Simionesco, \textit{J. Chem. Phys.} \textbf{117}, 2161 (2002). 
\bibitem{ROMpaper} T. Sarlat, A. Billoire, G. Biroli, J.-P. Bouchaud, {\it in preparation}.
\bibitem{gotze-sjogren} W. G\"otze, Sj\"orgen, \textit{Rep. Prog. Phys.} \textbf{55}, 241 (1992).
%\bibitem{vasiliev} A.N. Vasiliev Functional methods...
%\bibitem{zinn-justin} Zinn-Justin, \textit{Quantum Field Theory and Critical Phenomena}, Clarendon Press, Oxford.
%\bibitem{zwanzig} R. Zwanzig, \textit{Nonequilibrium Statistical Mechanics}, Oxford University Press, 2001.
\bibitem{WolynesIndians} S. M. Bhattacharya, B. Bagchi and P. G. Wolynes,  \textit{Phys. Rev. E} \textbf{72}, 031509 (2005). 
\bibitem{WE} For a related argument, see  M. P. Eastwood and P. G. Wolynes, \textit{Europhys. Lett.} \textbf{60}, 587-593 (2002).
\bibitem{VBG} M. Tarzia, G. Biroli, \textit{Europhys. Lett.} \textbf{82}, 67008 (2008). 
\end{thebibliography}
\end{document}